\documentclass[letterpaper, 10 pt, conference]{ieeeconf}
\usepackage{color}
\usepackage{framed}
\usepackage{amsmath}
\usepackage{amssymb}
\usepackage[noend]{algpseudocode}
\usepackage{amsfonts}
\usepackage[]{cite}
\usepackage{chngcntr}

\counterwithin{paragraph}{section}

\usepackage{amsthm}
\usepackage{etoolbox}
\newtoggle{showfig}
\newtoggle{RedundantLowerBound}
\newtoggle{RedundantText}

\togglefalse{showfig}
\togglefalse{RedundantLowerBound}
\togglefalse{RedundantText}

\usepackage[neverdecrease]{paralist}
\usepackage{graphicx}
	
\newtheoremstyle{plain}{\topsep}{\topsep}{\itshape}{}{\bfseries}{:}{.5em}{}%
\newtheorem*{theorem*}{Theorem}
\newtheorem*{lemma*}{Lemma}
\renewenvironment{proof}{{\bfseries Proof:}}{}
\IEEEoverridecommandlockouts                              
\begin{document}
\title{A General Framework for Pairs Trading with~a~Control-Theoretic~Point~of~View}
\author{Atul Deshpande$^{1}$ and B. Ross Barmish$^{2}$
\thanks{\hspace{-.25cm}${}^1$\hspace{1pt}Atul Deshpande is a graduate student working towards his doctoral dissertation in the Department of Electrical and Computer Engineering, University of Wisconsin, Madison, WI 53706.
        {\tt\small atul.deshpande@wisc.edu}}%
\thanks{\hspace{-.25cm}${}^2$\hspace{1pt}B. Ross Barmish is a faculty member in the Department of Electrical and Computer Engineering, University of Wisconsin, Madison, WI 53706.
        {\tt\small bob.barmish@wisc.edu}}%
}
\maketitle
\begin{abstract}
Pairs trading is a market-neutral strategy that exploits historical correlation between stocks to achieve statistical
arbitrage. \iftoggle{RedundantText}{Although commonly used in practice, literature on this topic is rather sparse, largely due to proprietary considerations. In fact, e}{E}xisting pairs-trading algorithms in the literature require rather restrictive assumptions on the underlying stochastic stock-price processes and the so-called {\it spread function}. In contrast to existing literature, we consider an algorithm for pairs trading which requires less restrictive assumptions than heretofore considered. Since our point of view is control-theoretic in nature, the analysis and results are straightforward to follow by a non-expert in finance. To this end, we describe a general pairs-trading algorithm which allows the user to define a rather arbitrary spread function which is used in a feedback context to modify the investment levels dynamically over time. When this function, in combination with the price process, satisfies a certain mean-reversion condition, we deem the stocks to be a tradeable pair.  For such a case, we prove that our control-inspired trading algorithm results in positive expected growth in account value. Finally, we describe tests of our algorithm on \iftoggle{RedundantText}{daily}{historical} trading data by fitting stock price pairs to a popular spread function used in literature\iftoggle{RedundantText}{. Simulation results on historical data}{}. Simulation results from these tests demonstrate robust growth while avoiding huge drawdowns.
\end{abstract}
%
\IEEEpeerreviewmaketitle
\section{Introduction}
A pairs-trading algorithm is a market-neutral strategy which exploits historical correlation between stocks to achieve statistical arbitrage. 
Such algorithms usually involve taking complementary positions in the two constituent stocks of the pair; i.e., long one stock and short the other. This occurs when the stock prices, which are otherwise historically related, temporarily diverge from their proven behavior. 
Under such conditions, the trader bets that the prices will move in a manner so as to return to their historical relationship.
Examples of correlated/paired stocks involve Exchange Traded
Funds (ETFs), certain currency pairs or stocks of companies in
the same industry such as Home Depot and Lowe’s or WalMart
and Target.\\[4pt]
Literature such as~\cite{Whi,Gatev2006,nath,DoSimple,DoWiley} deal with the more practical details of pairs trading and include considerations of the performance of pairs-trading methods, including the impact of transaction costs on profitability.
A \emph{co-integration model} between the logarithm of two stock prices is suggested in~\cite{Vid}, and the results are used to determine the magnitude of deviation of the spread from its equilibrium, which in turn triggers appropriate long/short positions on the pair.\\[4pt]
In~\cite{elliott}\iftoggle{RedundantText}{}{ and~\cite{Mudchanatongsuk2008}}, the logarithmic relationship between stock prices is modeled as an Ornstein-Uhlenbeck (O-U) process. \iftoggle{RedundantText}{This model is further used in~\cite{Mudchanatongsuk2008} to formulate a portfolio optimization-based stochastic control problem.}{}
The same model is used for the spread function of stock prices in the continuous-time setting in~\cite{Do2006}. However, the spread in this case is a function of stock returns and not the stock prices themselves.
Whereas the literature discussed so far deals with just one spread for a pair of stocks, reference~\cite{Kim2008} deals with multiple spreads, each involving baskets of underlying securities. Each of the aforementioned spreads is assumed to be modelled on an independent O-U process, and the paper proposes an optimal distribution of investment among the spreads.\\[4pt]
More recent papers such as~\cite{stochastic} and the {\it Cointelation} model in~\cite{cointelation} build up on the co-integration model of the spread function to design a stochastic control approach for pairs trading. 
To summarize, a vast preponderance of the theory developed to date requires rather specific assumptions on the underlying stock price processes and the spread function.\\[4pt]
{
Unlike the existing literature discussed above, the work described in this paper applies to not just one specific model of reversion for the spread as in ~\cite{Mudchanatongsuk2008,Do2006,Kim2008} and~\cite{cointelation}. Moreover, we do not limit the spread to be a particular function of the underlying stock prices as in~\cite{Vid, Mudchanatongsuk2008,Kim2008} and~\cite{stochastic} or make any assumptions on the underlying stock prices as in~\cite{stochastic}. We propose a general framework which works for an arbitrary spread function of the underlying stock prices, provided it satisfies the minimal requirements of mean-reversion as represented by an expectation condition given in the following section. Popular models like the Ornstein-Uhlenbeck process satisfy our conditions.}
\subsection{Control-Theoretic Point of View}
This paper falls within the recently emerging body of work on the application of control-theoretic concepts to stock trading; see~\cite{Barmish} and its bibliography for an overview of the relevant literature. At the same time, we note that little has been said about the application of control theory to pairs trading.
\begin{figure}[!t]
\centering
\includegraphics [width=3.5in]{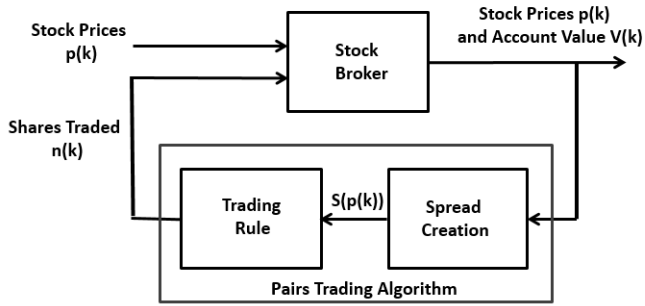}
\caption{Pairs Trading as a Stochastic Feedback Control Problem}
\label{pairsblock}
\vspace{-1.0em}\end{figure}
Our setup for the problem at hand is shown in Figure~\ref{pairsblock}. The stock price pair~$(p_1(k),p_2(k))$ is processed by the controller to create a spread function~$S(p(k))$ on which trading is based. The controller determines the number of shares~$(n_1(k),n_2(k))$ to be held in the respective stocks during each trading period. \\[4pt]
{One desirable property of our controller is to ensure positive expected change in account value at each step. In the following section, we state our assumptions regarding the market and the stocks which form a \emph{tradeable pair}. 
Given these assumptions, we provide a trading algorithm and prove that it results in positive expected growth in account value. 
We further test this algorithm on historical data by employing a spread function which is frequently used in literature, and see robust gains and low drawdowns in simulations.}
\section{General Framework for Pairs Trading}
\label{framework}
In this section, we state our assumptions of the market and the stock price processes. We further define a \emph{mean-reverting spread}, and provide a trading algorithm which guarantees positive expected growth in account value at each step when the investor holds positions in the stocks.\\[-15pt]
\subsection{Idealized Market Assumption}
The trader is assumed to be working under the following idealized market conditions. These requirements are nearly identical to those used in finance literature in the context of  
 ``frictionless markets'' dating back to~\cite{Merton} and used in many papers thereafter.\\[4pt]
{\bf Zero Transaction Costs}: The trader does not incur any transaction costs, such as brokerage commission, transaction fees or taxes for buying or selling shares.\\[4pt]
{\bf Price Taker}: The trader is a price taker who is small enough so as not to affect the prices of the stocks.\\[4pt]
{\bf Perfect Liquidity Conditions}: There is no gap between the bid/ask prices, and the trader can buy or sell any number, including \emph{fractions}, of shares at the currently traded price.\\[-7pt]
\subsubsection*{\bf Prices, Bounded Returns and Density Functions}
The stocks under consideration have strictly positive prices~$p_1(k)$ and $p_2(k)$. In addition, the price vector~$p(k)$ is assumed to have a continuous probability density function which is unknown to the trader.
For the stochastic stock-price process
$$
p(k) \doteq \left[\begin{array}{c}p_1(k)\\ p_2(k) \end{array}\right]
$$
with~$p_i(k)>0$ for~$i=1,2$, the return on the~$i$-th stock during the~$k$-th period is given by 
$$
X_i(k) \doteq \frac{p_i(k+1) -  p_i(k)}{p_i(k)}
$$
for~${i=1,2}$. Finally, we assume bounded returns. That is, there is a constant~${0<\gamma<1}$ such that~$
{|X_i(k)|\leq \gamma}
$~for~${i=1,2}$.
\subsection{Mean-Reverting Spread Function Assumption}
To motivate the definition to follow, we imagine a pair of stocks represented by the price process~$p(k)$ and view it as a {tradeable pair} if we can define a function~$S(p)$ such that the dynamics of the price processes cause~$S(p(k))$ to be \emph{mean-reverting} along sample paths. 
That is, the change in the spread function
$$
\Delta S(k) \doteq S(p(k+1)) - S(p(k))
$$
is expected to decrease the absolute value of the spread function. For example, when~$S(p(k))$ is positive along a sample path, we expect the price dynamics of its constituent stocks to force~$|S(p(k))|$ to reduce and move towards zero, with a ``pull'' proportional to its distance from zero.
In the formal definition to follow, this is manifested as the assumption that there exists some~$\eta>0$ such that when~${S(p(k))>0}$,
${
\mathbb{E}[\Delta S(k)|S(p(k))] \leq -\eta |S(p(k))|}
$.
Similarly, when~${S(p(k))<0}$, we expect symmetrical behaviour, that is,
$
\mathbb{E}[\Delta S(k)|S(p(k))] \geq \eta |S(p(k))|
$.\\[4pt]
In the above expression, the constant~$\eta$ is representative of the degree of reversion of the spread function. The expected ``pull'' is in a direction intended to reduce the spread, and the magnitude of the pull gets higher as the spread value increasingly deviates from zero.\\[-7pt]
\paragraph*{\bf Definition} 
A given twice continuously differentiable function ${S:(0,\infty)\times (0,\infty)\rightarrow \mathbb{R}}$ with no stationary points is said to be \emph{mean-reverting} with respect to the price process $p(k)$ if there exists a constant~$\eta>0$ such that the conditional expectation condition
$$
\mathbb{E}[\mbox{sign}(S(p(k)))\Delta S(k)|S(p(k))] \leq -\eta |S(p(k))|
$$
is satisfied.
\subsection{The Trading Algorithm}
With the above assumptions and definitions in place, we now describe an algorithm for trading the pair of stocks. To this end,
let~$V(k)$ be the value of the trading account at stage~$k$, with initial value~$V(0)$ and take~$\nabla S(p(k))$ to be the gradient vector of the spread function with respect to price~$p$ evaluated at~$p(k)$. Define~$|\nabla S(p(k))|$ as the vector of the absolute value of the elements of this vector. Let~$L>0$ be the allowed leveraging factor, that is, the trader is allowed to invest up to $L V(k)$ in absolute value.\\[-7pt]
\subsubsection*{\bf Trading Threshold}
We first describe the set of all possible values that~$p(k+1)$ can attain under the bounded returns assumption described in the previous section. Noting that
$$
\left|p_i(k+1) - p_i(k)\right| \leq \gamma p_i(k).
$$
and observing that $p(k+1)$ is contained in the set
$$
\mathcal{B}_\gamma(p(k))\doteq\{p:|p_i-p_i(k)|\leq \gamma p_i(k)\;\;\; \text{ for all } i\},
$$
we choose the trading threshold, a function of $p(k)$, to be
$$
\tau(k) \doteq \frac{1}{2\eta} \max_{p\in\mathcal{B}_\gamma(p(k))}\left|\frac{}{} (p-p(k))^T\nabla^2 S(p) (p-p(k))\right|
$$
where~$\nabla^2 S(p)$ is the Hessian of the spread function.\\[-7pt]
\subsubsection*{\bf Threshold-Based Trading Algorithm}
Let the vector~$n(k)$ represent the number of shares held by the trader in the stocks at the~$k$-th stage, with~$n_i(k)$ being the number of shares held in the~$i$-th stock. The following rule specifies the trader's holdings in the stocks:
\begin{samepage}
 Under favorable trading conditions, characterized by
${|S(k)|>\tau(k)}$, we take
$$
    n(k) = -\lambda(k) \text{\small sign}(S(p(k))) \nabla S(p(k))
$$
where 
$$
\lambda(k) \doteq \frac{LV(k)}{|\nabla S(p(k))|^T p(k)}
$$
is a positive constant.
Otherwise, we take
$$
n(k) = 0.
$$\\[-12pt]
\end{samepage}
Recalling that since $S$ has no stationary points, $\lambda(k)$ is always well defined.
Note that the \emph{idealized markets} allow us to trade fractional quantities of shares, a negative~$n_i(k)$ indicates \emph{shorting} that many shares of the~$i$-th stock, whereas a positive value indicates that the trader must \emph{buy} as many shares. On the other hand, $n(k)=0$ implies that the trader chooses not to hold any positions in the stocks.\\[4pt]
The formulae above guarantee that when~${|S(p(k))|>\tau(k)}$, the trader is fully invested up to the limit allowed by the broker, as total \emph{invested} amount
$
|n(k)|^T p(k) = L V(k).
$
For the special case when~$L=1$, the trader is said to be self financed.
At stage~$k+1$, the account value~$V(k+1)$ is redistributed in the stocks according to the trading algorithm described above, but with spread~${S(p(k+1))}$ as the input variable.\\[-10pt]
\paragraph*{\bf Remark} The change in account value during the~$k$-th interval is evaluated as
\begin{align*}
\Delta V(k) &= V(k+1) - V(k)\\[3pt]
&= n^T(k)\Delta p(k)
\end{align*}
where $\Delta p(k) \doteq p(k+1) - p(k)$ is the change in the price vector during the~$k$-th period.
In the following section, we  show that the trading algorithm described above yields positive expected growth in account value.\\[-10pt]
\section{Main Result}
According to our trading algorithm, for all $k$ such that~$S(p(k))\leq\tau(k)$, the trader does not hold any positions in the stocks, and hence
$
\Delta V(k) = 0.
$
The following theorem, the main result of the paper, tells us that when conditions are favorable for trading, the expected change in account value must be positive.\\[-12pt]
\begin{theorem*}[Positive Expected Growth]
Let~$(p_1(k),p_2(k))$ be a stock-price pair with bounded returns~$|X_i(k)|\leq \gamma$ for~$i = 1,2$ and associated spread function~$S(p)$ which is mean reverting with respect to sample paths~$p(k)$. Then the trading strategy with threshold~$\tau(k)$ guarantees that the expected change in the account value,~$\Delta V(k)$, is positive for all~$k$ for which trading occurs.
That is, for all $k$ such that
$P\left(|S(p(k))| > \tau(k)\right)>0,$
it follows that
$$
\mathbb{E}\left[\Delta V(k)\big||S(p(k))| > \tau(k)\right] > 0.
$$
\end{theorem*}
\subsection{Proof of the Postive Expected Growth Theorem}
We first state and prove a preliminary lemma which will later be used in the proof of the theorem:\\[-10pt]
\begin{samepage}
\begin{lemma*}[Bounded Approximation Error]
 Along the sample paths~$p(k)$, the difference between change in the spread function during the~$k$-th period and its linear approximation has bound
 {
 \begin{align*}
\left|\Delta S(k) - [\nabla S(p(k))]^T\Delta p(k)\right|
\leq \eta \tau(k).
\end{align*}
}
\end{lemma*} 
\end{samepage}
\begin{proof}
We consider the first-order Taylor series of the spread function for a given price change vector~$\Delta p$ from the price point~$p(k)$; i.e.,
$$
S(p(k)+\Delta p) = S(p(k)) + [\nabla S(p(k))]^T \Delta p + R_1(\Delta p)
$$
where the error term~$R_1(\Delta p)$ is the first-order Lagrange remainder.
In accordance with the Taylor-Lagrange formula~\cite{Taylor}, there exists a price point~$p^*$ and constant~${0<h^*<1}$,
such that with 
$$p^*= p(k)+h^*\Delta p,$$
it follows that
\begin{align*}
R_1(\Delta p) &= S(p(k)+\Delta p) - S(p(k)) - [\nabla S(p(k))]^T \Delta p \\[4pt]
&= \frac{1}{2} \Delta p^T \nabla^2 S(p^*) \Delta p.
\end{align*}
Recalling that
the change in the price vector~$\Delta p(k)$, although not known \emph{a priori}, is bounded, the range of possible values of~$p(k+1)$ is limited to the previously defined set $\mathcal{B}_\gamma(p(k))$.
The error term for this unknown~$\Delta p(k)$ is thus bounded~by
{\fontsize{9.5}{12}\selectfont
\begin{align*}
 |R_1(\Delta p (k))| &\leq \frac{1}{2} \max_{p\in\mathcal{B}_\gamma(p(k))} \left|(p-p(k))^T\nabla^2 S(p) (p-p(k))\right|\\[4pt]
 &= \eta \tau(k).
\end{align*}
}
Recalling the formula for 
$\Delta S(k)$
and the above bound on~$R_1(\Delta p(k))$, we obtain
\begin{align*}
&\left|\Delta S(k) - [\nabla S(p(k))]^T \Delta p(k)\right|\leq  \eta \tau(k). \qed
\end{align*}
\end{proof}
\paragraph*{\bf Proof of the Theorem}
Recalling that the change in account value
$$
{\Delta V(k) = n^T(k)\Delta p(k)}
$$
and substituting for~$n(k)$ from the definition in the previous section,  when $|S(p(k))|>\tau(k)$,
{
\begin{align*}
 \Delta V(k) = - \mbox{sign}(S(p(k)))\lambda(k)\left([\nabla S(p(k))]^T \Delta p(k)\right).
\end{align*}\\[-10pt]}
Since we are only interested  in proving that the sign of the expected change in account value is positive, and~$\lambda(k)>0$ is a constant for a given~$k$, without loss of generality, we assume~$\lambda(k)=1$. Thus,
\begin{align*}
 \Delta V(k) = - \mbox{sign}(S(p(k)))\left([\nabla S(p(k))]^T \Delta p(k)\right).
\end{align*}
From the \emph{Bounded Approximation Error} lemma, we identify the bounds
\iftoggle{RedundantText}
{
 {
 \begin{align*}
 [\nabla S(p(k))]^T \Delta p(k)
 \geq 
 \Delta S(k)
 -\eta \tau(k)
 \end{align*}}
 and
 {
 \begin{align*}
 [\nabla S(p(k))]^T \Delta p(k)
 \leq
 \Delta S(k)
+\eta \tau(k).
 \end{align*}}
}{
{ 
 \begin{align*}
 \Delta S(k)
-\eta \tau(k)\leq
 [\nabla S(p(k))]^T \Delta p(k)
 \leq
 \Delta S(k)
+\eta \tau(k).
 \end{align*}}
 }\\[-10pt]
 Using these inequalities, we obtain 
 {\fontsize{9.5}{12}\selectfont
 \begin{multline*}\hspace{-.42cm}
 \mbox{sign}(S(p(k)))
 [\nabla S(p(k))]^T \Delta p(k)
 \leq \mbox{sign}(S(p(k))) \Delta S(k)
 +\eta \tau(k).
 \end{multline*}\\[-10pt]
 }
Negating and taking expectation on both sides conditioned on~$S(p(k))$ leads to \iftoggle{RedundantLowerBound}{}{a lower bound for the expected change in account value conditioned on $S(p(k))$, namely}
{
\begin{align*}
 \mathbb{E}\left[\Delta V(k)\big|S(p(k))\right]&\\[4pt]
 &\hspace{-2.25cm}=-\mathbb{E}\left[\left. \mbox{sign}(S(p(k))) 
 \left([\nabla S(p(k))]^T \Delta p(k)\right)
 \right|S(p(k))\right]\\[4pt]
 &\hspace{-2.25cm}\geq - \mathbb{E}\left[\left. \mbox{sign}(S(p(k)))
 \Delta S(k)
 \right|S(p(k))\right] - \eta \tau(k).
\end{align*}\\[-10pt]}
\iftoggle{RedundantLowerBound}{
This gives a lower bound for the expected change in account value conditioned on $S(p(k))$, namely
 {
 \begin{align*}
  &\mathbb{E}\left[\Delta V(k)\big|S(p(k))\right]\\[4pt]
  &\hspace{1cm}\geq - \mathbb{E}\left[\left. \mbox{sign}(S(p(k)))
 \Delta S(k)
 \right|S(p(k))\right]
 - \eta \tau(k).
 \end{align*}\\[-10pt]
}}
Now invoking the mean-reversion assumption on~$S(p(k))$, we obtain
{
 \begin{align*}
  &\mathbb{E}\left[\Delta V(k)\big|S(p(k))\right]\\[4pt]
  &\hspace{.25cm}\geq -\mathbb{E}\left[\left. \mbox{sign}(S(p(k))) 
 \Delta S(k)
 \right|S(p(k))\right] - \eta \tau(k)\\[4pt]
 &\hspace{.25cm}>\eta\left(|S(p(k))| - \tau(k)\right).
 \end{align*}\\[-10pt]
}
Let~$f_{S(p(k))}(s)$ be the probability density function on~$S(p(k))$, perhaps discontinuous, induced by $p(k)$. Since~${P\left(|S(p(k))|>\tau(k)\right)>0}$,
the set 
$$
I_\tau(k) \doteq \{s: |s|>\tau(k) \text{ and } {f}_{S(p(k))}(s)> 0\}
$$
is non-empty with \emph{non-zero} length. Hence, noting that~${\mathbb{E}\left[\Delta V(k)\big|S(p(k))\right]>0}$ for all~${S(p(k))\in I_\tau(k)}$ and using the \emph{Law of Total Expectation}, we obtain
\begin{align*}
 \mathbb{E}\left[\Delta V(k)\big||S(p(k))|>\tau(k)\right]\\[4pt]
 &\hspace{-3cm}= \int\limits_{s\in I_\tau(k)}\mathbb{E}\left[\Delta V(k)\big|S(p(k))\right]{f}_{S(p(k))}(s) ds>0.\qed
\end{align*}
\section{Simulations and Results}
In this section, we first describe our general simulation setup. Then we use a candidate pair of securities and spread function, and simulate our trading algorithm using historical data. Finally, we present and discuss the results and compare the performance of our algorithm to buy-and-hold strategies on the constituent securities of the pair.
All the simulations to follow use the leveraging factor~$L=1$; that is, we assume a self-financed account. Additionally, the algorithm ensures that the trader is fully leveraged whenever he takes positions in the stocks.
\subsection{Simulation Setup}
To test our algorithm on historical data, we first select candidate securities for pairs trading and a candidate spread function. Then, via use of historical data, we fit the security prices to the candidate spread function, and check for statistical satisfaction of the {mean-reversion} property.\\[4pt]
Once trading begins, using the data withheld, we use a staggered sliding window method to estimate model parameters on the fly. This departure from strict application of the theory is done because the relationship between the stock prices is not necessarily stationary in practice. In this framework, we use \emph{training windows} of length~$N$, followed by \emph{trading windows} of length~$m$. At the end of the training window, current model parameters specific to the spread function are estimated. Then, this model is used to calculate the spread function and the threshold during the trading window which immediately follows. 
In our simulations, we use~$N=40$ and~$m=5$.\\[4pt]
During the training window, we also calculate the returns using the prices of the securities. The maximum absolute value of these returns leads to our estimate~$\hat{\gamma}(k)$, namely,
$$
\hat{\gamma}(k) \doteq \max_{i = 1,2;\; k-N\leq j < k-1
} \left|X_i(j)\right|.
$$
Then, we use the spread function computed over the training window in the previous step in conjunction with a sample-average derivative of the mean-reversion condition to obtain the estimate 
 \begin{align*}
 \hat{\eta} &\doteq - \frac{\sum\limits_{j=k-N}^{k-2} {\text{sign}(S(j))\left(S(j+1)-S(j)\right)}}{\sum\limits_{j=k-N}^{k-2}|S(j)|}.
\end{align*}
The formula above implicitly deweights samples for which~$S(p(k))$ is very small, so that the high relative change with respect to those does not impact the estimation process.\\[4pt]
We also use our knowledge of the spread function model to compute the Hessian~${\nabla^2 S(p)}$ at~$p(k)$.
Using the parameters esimated above, we compute the treshold as
{\fontsize{9.1}{12}\selectfont
\begin{equation*}
 \hspace{-.23cm}\hat{\tau}(k) \doteq \begin{cases}
\frac{1}{2 \hat{\eta}}\max\limits_{p\in\mathcal{B}_\gamma(p(k))}\left|(p-p(k))^T\nabla^2 S(p) (p-p(k))\right| &\text{ if }\hat{\eta}>0;\\
\infty &\text{ if } \hat{\eta}\leq 0
\end{cases}
\end{equation*}
}\\[-5pt]
where 
$
\mathcal{B}_\gamma(p(k))
$
is as defined in the previous section.
During the trading window, we evaluate the spread function and compare its magnitude with the threshold calculated above, and if 
$
|S(p(k))|>\tau(k),
$
we hold $n_1(k)$ and $n_2(k)$ shares, calculated according to the trading algorithm described in the previous section.
\subsection{Example - YINN and YANG}
\begin{figure}[!t]
\centering
\includegraphics [width=3.5in]{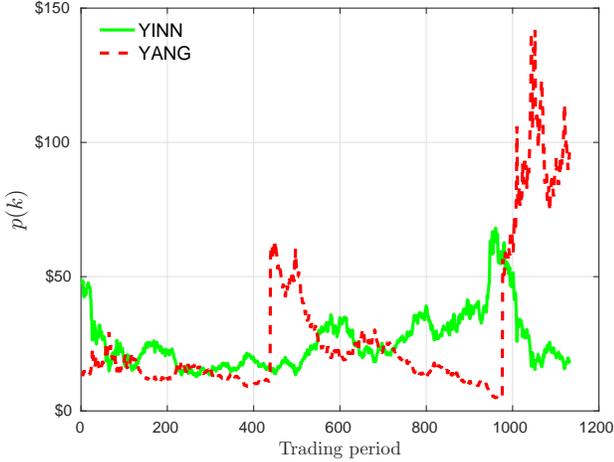}
\caption{Daily Closing Prices of YINN and YANG}
\label{yinnyang}
\vspace{-1.0em}\end{figure}

\begin{figure}[!t]
\centering
\includegraphics [width=3.5in]{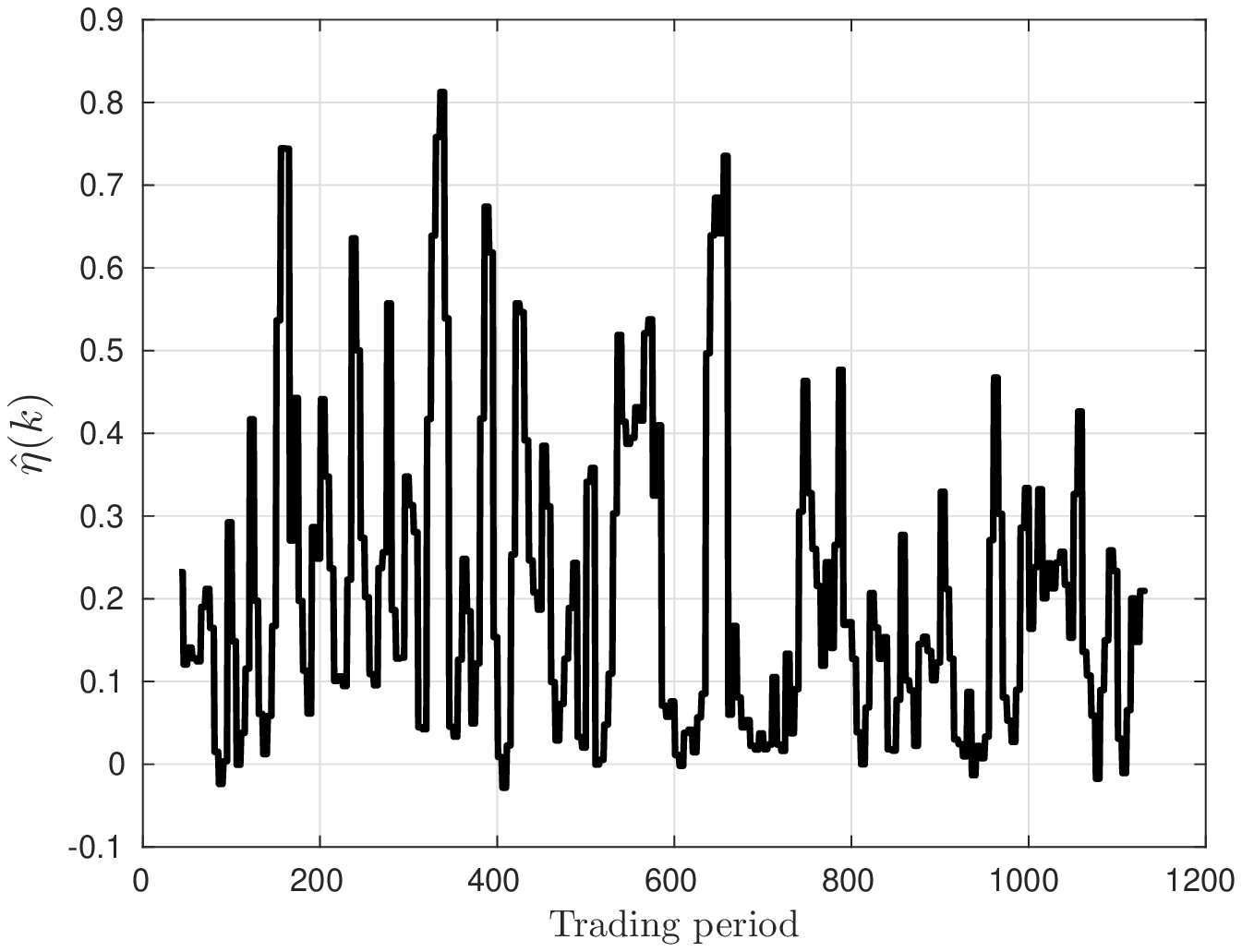}
\caption{$\hat{\eta}(k)$ Estimated from the Spread Function}
\label{eta}
\vspace{-1.0em}\end{figure}

The pair of securities chosen for testing were the exchange-traded funds Direxion Daily FTSE China Bull 3X ETF (YINN) and the Direxion Daily FTSE China Bear 3X ETF (YANG). These are related to the same market, namely China, albeit with different outlooks. Also, since both the ETFs are 3X leveraged in the markets, they are more volatile, leading to more frequent trading opportunities.
Figure~\ref{yinnyang} shows the daily closing prices of these two securities for the period from~July~1,~2011 to~December~31,~2015. Noting the price corrections made by the fund management in YANG around trading periods~$438$ and~$936$ respectively, we correspondingly adjust these prices before using them for analysis.\\[4pt]
First, we select the co-integration model used in prior literature for the spread function; namely
 \begin{align*}
 S(p) = log(p_2) - \beta log(p_1) -\mu.
\end{align*}
 Once trading has commenced, we fit the price data to our chosen model using a regression on $S$ above to obtain the estimates~$\hat{\beta}$ and~$\hat{\mu}$ during the training window. \iftoggle{showfig}{Figure~\ref{beta_mu} shows the evolution of these estimates over time, underlining the changing price relationship between the two securities.}\\[4pt]
 Finally, using this model, we compute the spread function retrospectively over the training window, and also use it during the trading~window.
 \iftoggle{showfig}{
\begin{figure}[!t]
\centering
\includegraphics [width=3.5in]{beta.eps}
\includegraphics [width=3.5in]{mu.eps}
\caption{$\beta(k)$ and~$\mu(k)$ Estimated via Regression}
\label{beta_mu}
\vspace{-1.0em}\end{figure}}
We use the constructed spread function over the training window to estimate $\hat{\eta}$ as described before. Figure~\ref{eta} shows the estimated~$\hat{\eta}(k)$ versus trading period.
We note that a near-zero or negative~$\hat{\eta}(k)$ is interpreted as unfavorable conditions for pairs trading. That is, the requirement $|S(p(k))|>\tau(k)$ becomes nearly impossible to satisfy.\\[4pt]
\begin{figure}[!t]
\centering
\includegraphics [width=3.5in]{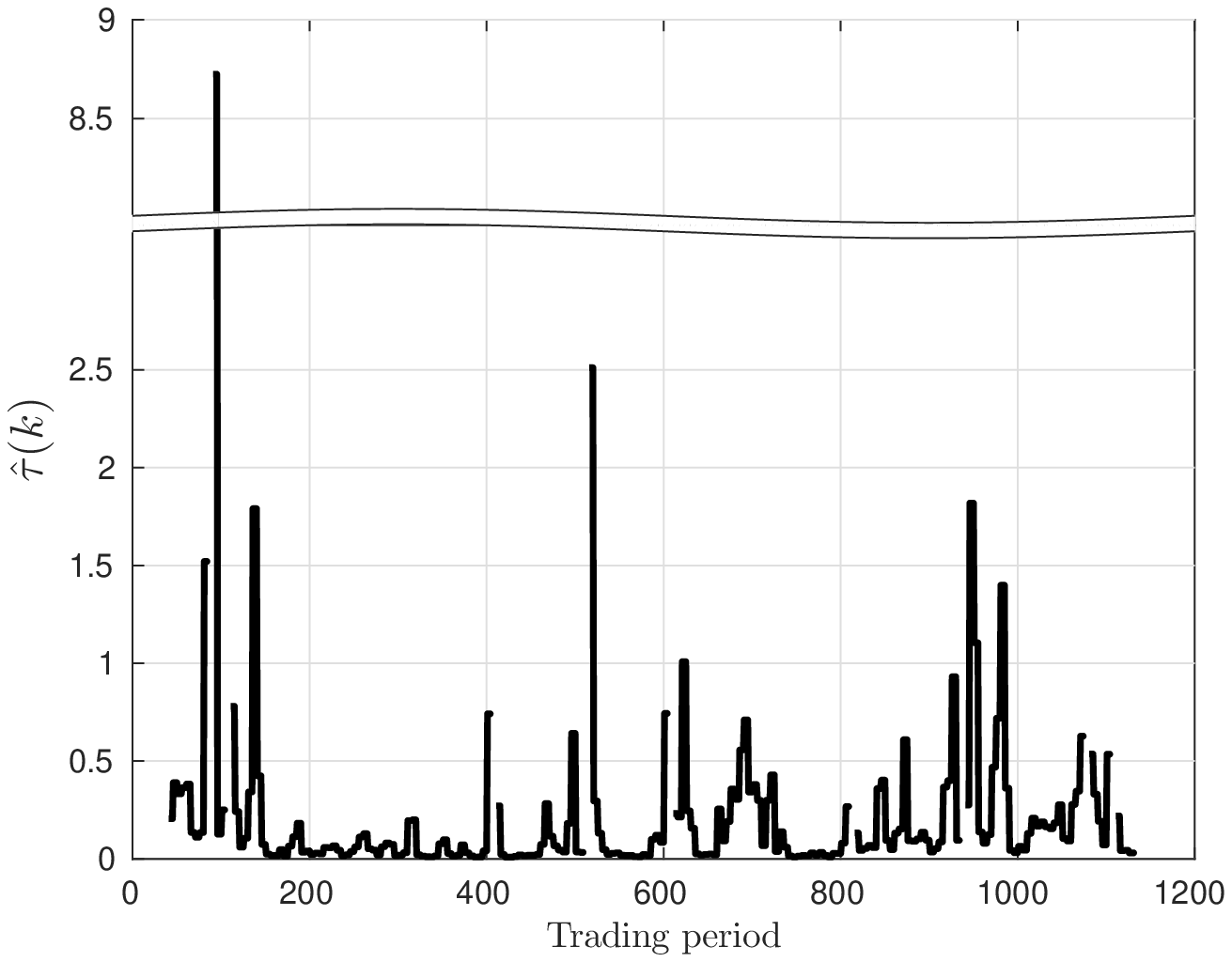}
\caption{$\hat{\tau}(k)$ Calculated Using~$\hat{\eta}$,~$\hat{\gamma}(k)$ and~$\nabla^2 S(p)$}
\label{tau}
\vspace{-.0em}\end{figure}\\[-10pt]
We now use our knowledge of~$\hat{\beta}$ to compute an estimate of the Hessian~${\nabla^2 S(p(k))}$ using the formula
~$$
 \nabla^2 \hat{S}(p(k)) \doteq \left[
 \begin{array}{cc}
  -\frac{\hat{\beta}(k)}{p_1^2(k)}&0\\
  0&\frac{1}{p_2^2(k)}
 \end{array}
 \right].
~$$
The running estimate~$\hat{\eta}(k)$ and~$\nabla^2 \hat{S}(p(k))$ are used to compute $\hat{\tau}(k)$. For simplicity of computation, in the calculations to follow, we approximate the Hessian as a constant over~$\mathcal{B}_\gamma(p(k))$ and work with the estimate
\begin{equation*}
 \hat{\tau}(k) = \begin{cases}
\frac{\hat{\gamma}^2}{2 \hat{\eta}}\left|p(k)^T\nabla^2 \hat{S}(p(k)) p(k)\right| &\text{ if }\hat{\eta}>0\\
\infty &\text{ if } \hat{\eta}\leq 0
\end{cases}
\end{equation*}
Figure~\ref{tau} shows the trend for~$\hat{\tau}(k)$ over time. 
The y-axis is broken to better represent the variation in the lower values of $\hat{\tau}(k)$ while simultaneously capturing the occasional high value.
Note that the plot of~$\hat{\tau}(k)$ is discontinuous in $k$, and the breaks indicate times when~$\hat{\tau}(k)=\infty$; this occurs when~$\hat{\eta}\leq 0$. The values of the computed spread function and the $\hat{\tau}(k)$ are compared to determine whether conditions are favorable for trading, and if they are, the share holdings are determined in accordance with the trading rule presented in the previous section.\\[-5pt]
\subsubsection*{\bf Results}
\begin{figure}[!t]
\centering
\includegraphics [width=3.5in]{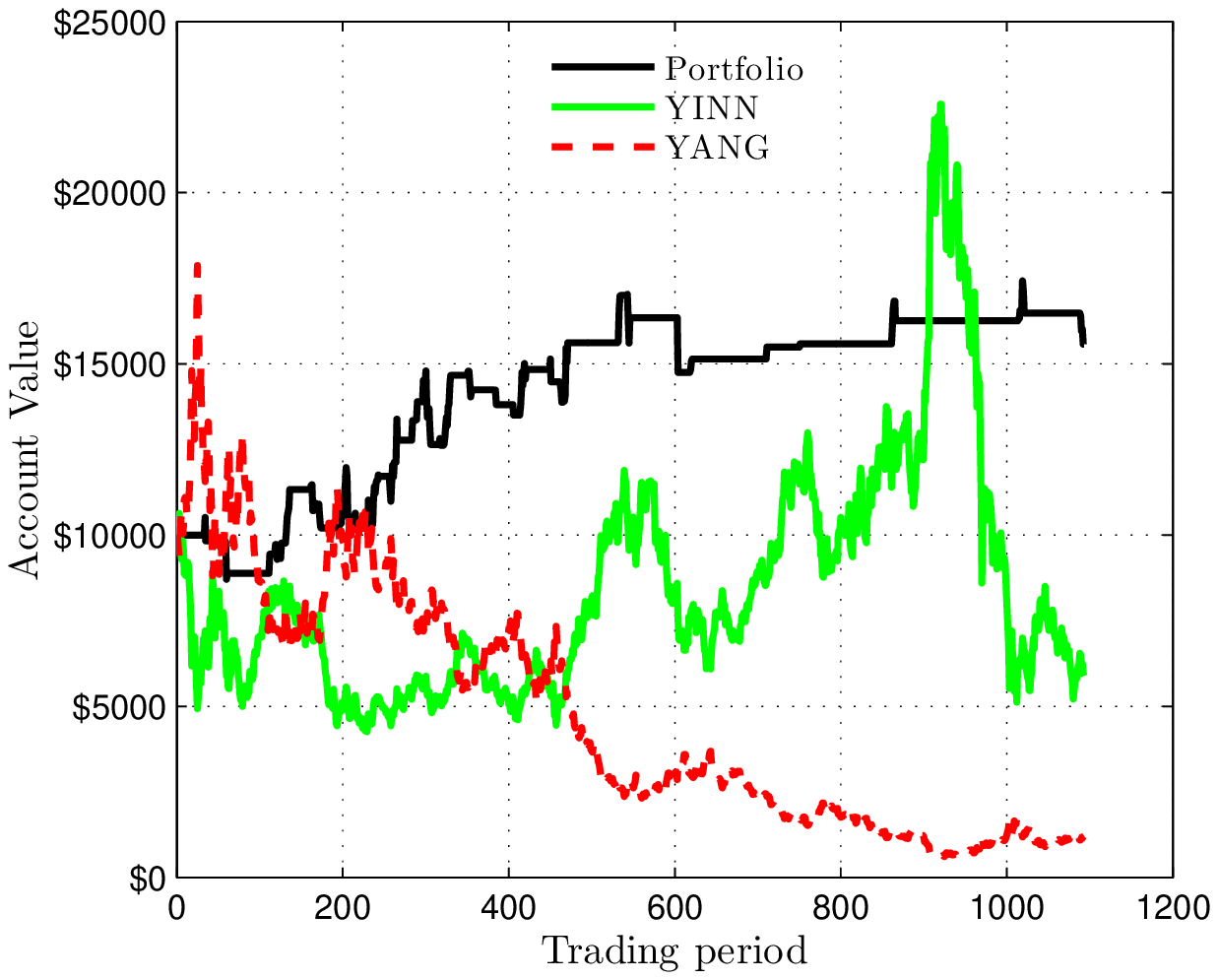}
\caption{Pairs Trading Compared to Performance of YINN and YANG}
\label{theorytau}
\vspace{-1.0em}\end{figure}

\begin{samepage}
To evaluate the performance of our trading algorithm, we consider three separate scenarios.
The first two of these correspond to a straightforward buy-and-hold beginning with \$10,000 worth of YINN securities and \$10,000 worth of YANG securities respectively.
The third scenario corresponds to using our threshold-based algorithm to trade the two securities with a starting account value of \$10,000.
\end{samepage}\\[4pt]
Figure~\ref{theorytau} shows the performance of the three scenarios during the period under consideration. As seen in the figure, a trader invested solely in YANG initially sees a 78\% profit, but eventually loses nearly 88\% of the account value. On the other hand, a trader invested solely in YINN loses 41\% of the account value after seeing a peak profit of 126\%. 
By design, these securities are bullish and bearish respectively on the same index, and in an ideal world, one would expect the losses in one portfolio to be offset by profits in the other. But as seen from Figure~\ref{theorytau}, both these scenarios eventually turn out to be loss-making.  This can be explained by the fact that the ETFs often fail to accurately track their target indices, and the operation of leveraged ETFs comes with additional risks and overheads, as explained in \cite{ETF}.\\[4pt]
The portfolio which trades using our algorithm shows 60\% profits over the same period. It is also noteworthy that despite the high volatility in the securities, the pairs-trading strategy results in minimal drawdowns. The coincidence of a majority of the gains shown by the portfolio with the periods when $\hat{\eta}(k)$ achieves high values in Figure~\ref{eta} points to the potential for future work involving the efficacy of $\eta$ as an indicator of fit quality between a spread function and the price data: see Section~5 for further discussion.
Also, during the worst period for the YINN portfolio (900-1000), the pairs never trade as a result of a high~$\hat{\tau}(k)$. 
This suggests a possible explanation as to why we avoid the disastrous drawdowns which wiped out the gains in the buy-and-hold trading scenarios used for comparison.\\[-5pt]
\vspace{-5pt}
\section{Conclusion}
In this paper, we presented an algorithm for
trading a pair of securities under rather weak hypotheses on the 
the price process and the spread function being used. Under the assumptions of bounded returns and mean-reversion described in Section~2, we described a threshold-based trading scheme which guarantees positive expected growth in the account value. To illustrate how the trading algorithm works in practice, we provided simulation results involving a pair of exchange-traded funds. Our results show robust growth compared to an alternative  buy-and-hold strategy which one might use on the constituent securities.\\[4pt]
The first point to note is that the theory presented in this paper includes three assumptions which were made solely for the purpose of simplifying the exposition.  First, we assumed that only two stocks are involved in the spread. In fact, if we consider a spread which is comprised of more than two stocks, the analysis of the account value is nearly identical to that given here. This type of more general portfolio-like problem will be pursued in our future work. The second assumption we made is that each price~$p_i(k)$ is a random variable with a continuous probability density function. In fact, the proof of the main theorem can easily be extended to handle the case when only a probability measure is available. Finally, we assumed that the stocks have bounded returns. However, even when these assumptions are dropped, we believe it should be possible to analyze the case when the returns are bounded with an appropriately high probability, and obtain similar results.\\[4pt]
By way of future research, further study of the estimated mean-reversion parameter~$\hat{\eta}(k)$ seems promising. Given a pair of securities, by observing this variable using training data, it would be of interest to study the extent to which~$\hat{\eta}(k)$ is a predictor as to the ``promise'' of a pairs trade. A second topic for future study is that of trading frequency. Given that our simulations were carried out using daily closing prices, it would be of interest to see how our algorithm performs when prices arrive more frequently. Studies of this nature should be possible to carry out using available tick data. More generally, there may be a number if important optimization problems associated with the issues raised above and our approach to pairs-trading problems. From a practical perspective, it would be of interest to include a number of considerations such as margin, risk-free securities and transaction costs in future analyses.
\bibliographystyle{IEEEtran}




%



\end{document}